\documentclass[aps,twocolumn,prl,unsortedaddress]{revtex4}%
\usepackage{amsmath}
\usepackage{graphicx}%
\usepackage{amsfonts}%
\usepackage{amssymb}
\begin{document}

\title{Numerical Test of the Energy Scale of Columnar Dimerization in High-T$_{c}$
Cuperate Superconductors}
\author{Oron Zachar}
\affiliation{UCLA Dept. of Physics, Los Angeles, CA 90095, USA.}

\begin{abstract}
Spontaneous bond dimerization (BD) was cojectured to be central to the physics
of the cuperate superconductors. Based on ladder models and numerical
simulations, we examine the quantitative issue of whether the energy scales
involved are sufficient to influence the mechanisms of pairing, stripe
formation, pseudo-gap, and quantum critical `normal' state phenomena. We
present preliminary evidence that maximum gap scales derived from BD are
always too small.

\end{abstract}
\maketitle

A mechanism of spontaneous bond dimerization (BD) was proposed
\cite{Sachdev2000-Science,Sachdev01-PhaseDiagram,SachdevVojta2000-PhaseDiagram,SachdevRead90-BondOrder}
as an underlying core phenomenon in the cuperate high temperature
superconducting materials. Columnar BD was proposed as the mechanism for the
formation of stripes in underdoped cuperates (see Fig.1). A quantum critical
point associated with BD is proposed to be responsible for certain anomalous
thermodynamic and magnetic properties observed in the temperature-doping phase
diagram of the cuperates. Elaborate consequences for impurity and vortex
states were derived
\cite{Sachdev01-PhaseDiagram,SachdevVojta2000-PhaseDiagram}. Moreover, the BD
is characterized by a spin gap and associated confinement of spin-$\frac{1}%
{2}$ excitations, and hence we note that by itself it also provides a
potential mechanism of pairing.

The analysis in Refs.
\cite{Sachdev2000-Science,Sachdev01-PhaseDiagram,SachdevVojta2000-PhaseDiagram}
focussed primarily on symmetry aspects of the BCS and BD order parameters.
Here we focus on quantitative issues. In the cuperates, the maximum spin-gap
in the superconducting state is $\Delta_{s}\equiv2\Delta\left(  0\right)
\approx800K\sim J/2$ \cite{PseudoGap-review-Timusk99}, where $J$ is the
antiferromagnetic exchange interaction. Pseudo-gap phenomena persist to
temperatures of order $300K$. Some evidence of local stripe order
(incommensurate magnetism) persists up to temperatures of order $70K$ in some
underdoped materials, where spectral signatures of incommensurate magnetism
are seen up to energies of order $40meV$ ($\sim450K$) \cite{StripesEnergy}.
Thus, for these phenomena to be associated with BD order or fluctuations, this
order must generate an energy scale, say a spin gap (or spin pseudo gap),
which is sufficiently large. We will adopt the lenient criterion that this
energy scale must be at least of order $\Delta>250K\approx J/6$.

\emph{Our main observation is that this aspect of the BD conjecture can be
tested by numerical simulations on wide ladders} (e.g., 6-leg or 8-leg).
Indeed, existing numerical experiments on 4-leg ladders and on small square
Heisenberg arrays will be shown to already suggest rather strongly that the
energy scales derived directly from BD are considerably smaller than $250K$.
Pending sharper similar results in 6-leg ladders, we doubt the viability of BD
as a candidate theory of pairing or stripes in the cuperate high-$T_{c}$ superconductors.

We stress that our analysis bares strictly on the mechanism of spontaneous BD.
Associated results of the work by Sachdev and collaborators may be relevant to
more general models of stripes or ladders, and hence their relevance to the
cuperate superconductors may hitherto not be tied to the mechanism of
spontaneous BD.%

\begin{figure}
[ptb]
\begin{center}
\includegraphics[
height=2.5832in,
width=2.7371in
]%
{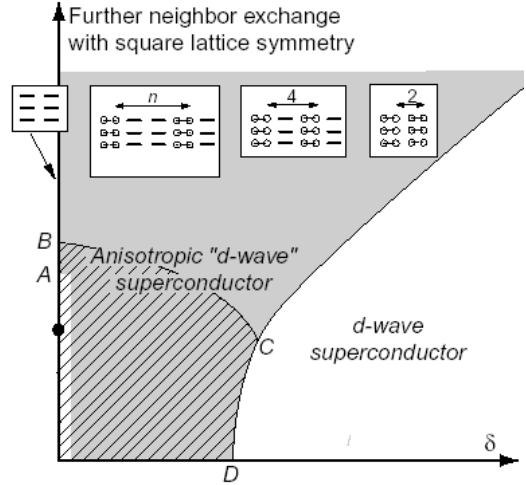}%
\caption{Conjectured phase diagram by Sachdev et \textsl{al.}
\cite{Sachdev01-PhaseDiagram}.}%
\label{Fig-SachdevPhaseDiagram}%
\end{center}
\end{figure}

\section{Review of the columnar dimerization approach to High-Tc}

Columnar BD was introduced as a bond ordered state in 2D Heisenberg models
with frustrated interactions \cite{SachdevRead90-BondOrder}. The canonical
example is the model with antiferromagnetic nearest-neighbor and
next-nearest-neighbor interactions ($J$ and $J^{\prime}$ respectively). The
core physics can be generalized to the inclusion of further neighbor
interactions. The bond ordered state is proposed to correspond to the spin
gaped state which occurs when $0.4<J/J^{\prime}<0.6$
\cite{ChakravartyHalperinNelson89,Chandra88-J1J2model,Kivelson89-J1J2RVB,Capriotti2002-J1J2Heisenberg}%
. The columnar BD order renders the system to be equivalent to an effective
coupled 2-leg spin ladder system (see Fig.2). It is supposed that at moderate
doping the BD and spin gap are retained
\cite{Sachdev2000-Science,Sachdev01-PhaseDiagram} which entails the
confinement of spin-$\frac{1}{2}$ excitations (i.e., pairing). The spin gap is
the prominent energy scale which characterizes the BD. In the spirit of the
RVB doped spin liquid idea\cite{RVB}, pairing is inherited from the spin
liquid state of the undoped BD Heisenberg model. An analogy is drawn with the
case of dimerization in one dimensional spin chains, where the spin gap can be
as large as $J$ (the Majumdar-Ghosh limit). A-priori, it might be that
columnar BD can lead to a spin gap on the scale of weakly coupled 2-leg
ladders \cite{Sachdev2000-Science,Sachdev01-PhaseDiagram}.%

\begin{figure}
[ptb]
\begin{center}
\includegraphics[
height=0.9703in,
width=3.0277in
]%
{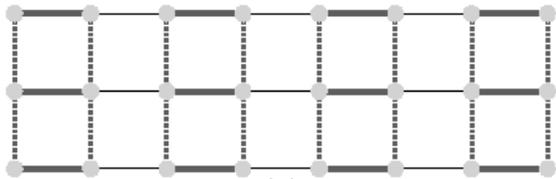}%
\caption{Columnar bond dimerization (BD): It is conjectured by Sachdev et
\textsl{al.}\cite{Sachdev01-PhaseDiagram,SachdevVojta2000-PhaseDiagram} that
the low energy physics is governed by an effective Hamiltonian of a 2D $t$-$J$
model where the Heisenberg interaction $J$ is alternating in strength along
one axis.}%
\label{Fig-ColumnarDimerization}%
\end{center}
\end{figure}

Since the undoped cuperates have a quasi long range antiferromagnetic ground
state \cite{ChakravartyHalperinNelson89,Aeppli2000-LCO-SpinWaves}, the
applicability of a BD model to the cuperates is not straight foreword. The
mechanism implied by the phase diagram (Fig.1)\cite{Sachdev01-PhaseDiagram} is
the following. The microscopic degrees of freedom and interactions in the
cuperate superconductors are capture by a one band generalized t-J model
(e.g., t--t'-J-J'-J'') plus coulomb interactions. The \emph{bare parameters}
values are such that groundstate of the undoped system is the Neel state as
described by Chakravarty \textsl{et al. }\cite{ChakravartyHalperinNelson89}.
With the introduction of holes, the holes dynamics enhances the effective
longer range antiferromagnetic interactions (i.e., imagine integrating out
high energy hole hopping processes to arrive at an effective lower energy
Hamiltonian). Thus, at intermediate energy scales, it is conjectured that the
\emph{effective spin interactions model parameters} are in the fixed point
regime of the bond dimerized Heisenberg model. The low energy holes dynamics
is self consistently centered on the strong bonds (which in columnar BD form
seem very much like 2-leg ladders).

The addition of coulomb interaction has the effect of changing the spacing of
the hole rich columns compared with the model without such coulomb interaction
\cite{SachdevVojta2000-PhaseDiagram}. Hence, a stripe like structure of
alternating hole rich lines and hole poor domains can be formed (see insets in
Fig.1). We comment that the BD as envisaged in
\cite{Sachdev01-PhaseDiagram,SachdevVojta2000-PhaseDiagram} is a property not
only of the hole rich stripe but also of the hole poor environment between
stripes. Thus, the spin gap in the hole rich stripe domain is tightly linked
with properties of the hole poor domain between stripes.

\section{Critical test of the prospects of columnar dimerization}

Prior to discussing direct tests of the spontaneous BD mechanism, one may ask
what are its most idealized prospects to begin with? Numerical simulations
results of for the spin gap on a square $J_{1}-J_{2}$ Heisenberg model are
shown in Fig-3 \cite{Capriotti2002-J1J2Heisenberg}. A spin gap is expected
only in the range of parameters $0.4<J_{2}/J_{1}<0.6$. It appears that, unlike
the one dimensional spin chain case where the spin gap can be as big as
$\Delta_{s}\approx J_{1}$ (the Majumdar-Ghosh limit), the effective
\emph{maximal} dimerization in a 2D $J_{1}-J_{2}$ Heisenberg model on a square
lattice is never giving rise to a gap value bigger than $J_{1}/8$. While it
may be that by fine tuning some additional further interactions ($J_{3}%
,J_{4},$ etc...) one might conjure higher gap values, that would not
constitute a robust model. Since the square $J_{1}-J_{2}$ Heisenberg model is
the effective model system which is at the origin of the BD mechanism proposed
by Sachdev, the results presented in Fig.3 already put in serious doubt the
basic conjecture which underlies the introduction of spontaneous BD as
responsible for phenomena at energy scale $\Delta\gtrsim J/6$ in the cuperate
high temperature superconductors.%

\begin{figure}
[ptb]
\begin{center}
\includegraphics[
height=2.0003in,
width=3in
]%
{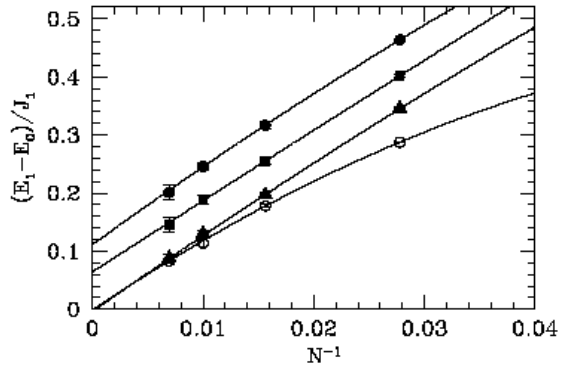}%
\caption{Numerical simulation results for the spin gap of $J_{1}-J_{2}$
Heisenberg models\cite{Capriotti2002-J1J2Heisenberg}. Size scaling of the
energy gap to the first $S=1$ spin excitation. Largest system is $N=144$.
$J_{2}/J_{1}=0.38$ (triangles), $0.45$ (squares), $0.50$ (full circles). For
comparison, the unfrustrated ($J^{\prime}=0$) Heisenberg model data is given
(empty circles).}%
\label{Fig-J1J2-gap}%
\end{center}
\end{figure}

Even leg ladders are an ideal testing ground for t-J model physics. Numerical
simulations of doped t-J model ladders
\cite{White2002-Doped4leg,Poilblanc02-4leg-Gap} show groundstate pairing
correlations in agreement with effective columnar BD (see Fig.4). Therefore,
ladders seem to be a favorable system for BD. It is worth noting that the
appearance of columnar pairing correlations in ladders may also be a simple
consequence of geometrical anisotropy and boundary conditions. Yet, such
knowledge is not essential to our arguments in this paper. We argue that
\emph{the key issue is the size of the spin gap}.%

\begin{figure}
[ptb]
\begin{center}
\includegraphics[
height=1.107in,
width=2.7657in
]%
{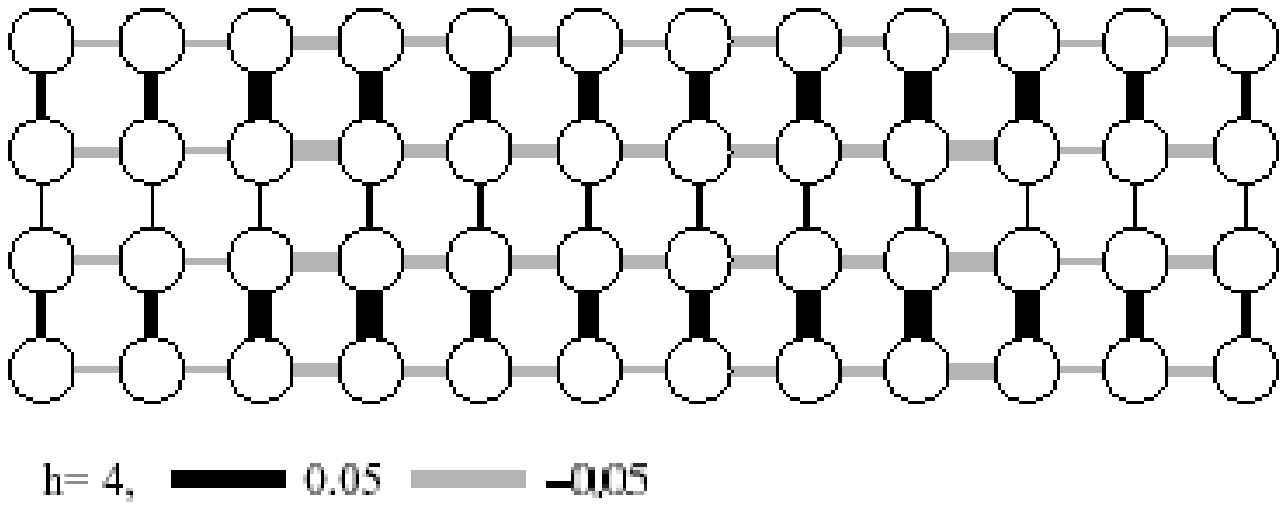}%
\caption{DMRG results for the real space singlet pairing expectation values on
nearest-neighbor sites\cite{White2002-Doped4leg}. Pair correlations on doped
4-leg ladder seem to manifest columnar dimerization. }%
\label{Fig-LadderPairCorrelations}%
\end{center}
\end{figure}

Numerical simulations by CORE method on doped 4-leg ladders already seem to
indicate that the spin gap of the doped ladder does not get bigger than that
of the undoped ladder (see Fig.5). But there is an ambiguity between open and
periodic boundary conditions. Apparently, the spin gap of 4-leg ladders
($\Delta_{0}\approx J/5$) is to begin with too large to draw sharp conclusions
from. On the other hand, as we elaborate below, similar calculations for 6-leg
ladders are all that is needed to make the point.%

\begin{figure}
[ptb]
\begin{center}
\includegraphics[
height=1.7339in,
width=2.9386in
]%
{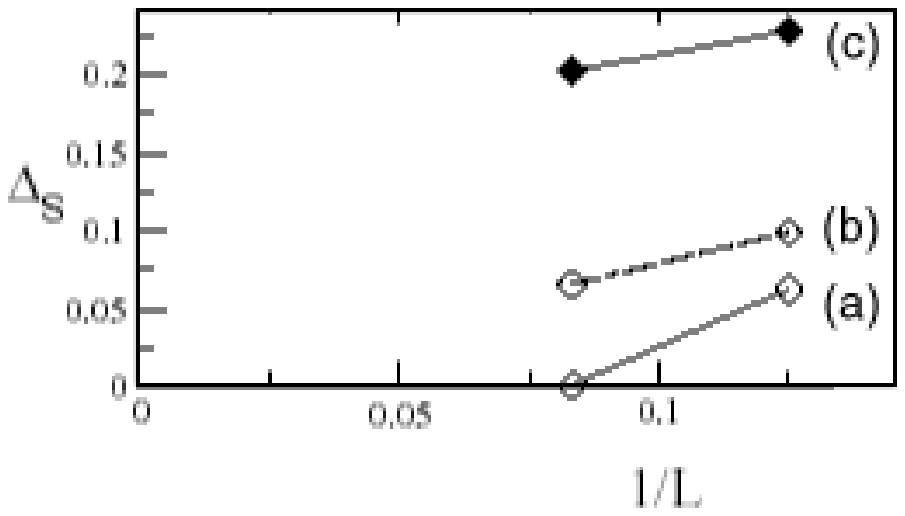}%
\caption{DMRG results for the real space singlet pairing expectation values on
nearest-neighbor sites\cite{White2002-Doped4leg}. Pair correlations on doped
4-leg ladder seem to manifest columnar dimerization. }%
\label{Fig-4leg-Gap}%
\end{center}
\end{figure}

Undoped 6-leg t-J ladder models have a small spin gap $\Delta_{0}\approx
J/20$. Upon doping, the ground state still has a spin gap and enhanced pairing
correlations. We propose to examine a 6-leg ladder model with the same bare
t-J-J' model parameters as are assumed for the C-O planes of the cuperate
superconductors (e.g., bare ring exchange interaction may need to be included
\cite{Aeppli2000-LCO-SpinWaves}). The generation of effective interactions due
to the holes dynamics is a local process which qualitatively and
quantitatively we expect to be the same in wide ladders (e.g., 6-leg or 8-leg
ladders) an in a 2D plane. Consequently, what ever is the potential for BD in
the 2D C-O plane, it should manifest itself with the same strength already in
wide ladder models.

The spin gap in undoped and doped 6-leg $t-J$ ladders can be deduced from the
energy gap to the first $S=1$ spin excitation. Examples of such a calculation
were done for a 2-leg and 4-leg ladders by both DMRG and CORE methods
\cite{Poilblnanc2000-doped2leg-Gap,Poilblanc02-4leg-Gap}. For spontaneous BD
to be a mechanism relevant at energy scales of order $\Delta\approx250K$, it
is required that the size of the spin gap for the doped system will be on the
order of $\Delta_{s}\approx250K$, i.e., 4 times bigger than that of the
undoped 6-leg ladder!!

In summary, bond dimerization (BD) is seen to occur, to some extent, in doped
even-leg ladders. Thus, aspects of the resulting physics can be directly
studied by numerical investigations of such ladders.
Specifically, we have proposed testing whether the energy scales involved are
sufficiently large to be relevant to a host of properties of the cuperates
(such as pairing and stripes).
The experimental phenomenology is sufficiently complicated that detailed
quantitative comparisons are certainly premature, but we believe there is a
significant theoretical distinction between the \emph{maximum} spin-gap energy
scales $\Delta_{s}\approx J/8$ found in BD
magnets\cite{Capriotti2002-J1J2Heisenberg}, and the \emph{minimum necessary}
magnitudes $\Delta_{s}>J/6$. More directly, if the BD mechanism fails to
produce a high spin gap energy scale in doped 6-leg and 8-leg ladder systems
(as is our prediction) then its prospects as the root cause of pairing, stripe
formation, or normal state properties in the cuperate superconductors are in
serious doubt. Of course, we cannot rule out the possibility that additional
well adjusted interactions, such as the electron-phonon interaction, could
increase the stability of the BD state and hence manage to overcome this
objection. Yet, such elements are not part of the framework currently
presented by
\cite{Sachdev2000-Science,Sachdev01-PhaseDiagram,SachdevVojta2000-PhaseDiagram,SachdevRead90-BondOrder}%
.

\begin{acknowledgments}
We thank Luca Capriotti, Doug Scallapino, and Steve Kivelson for illuminating
discussions. This work was supported by NSF grant \#4-444025-KI-21420-2.
\end{acknowledgments}

\end{document}